\documentstyle[aas2pp4]{article}

\newcommand{\dt}[1]{\frac{\partial #1}{\partial t}}
\newcommand{\Dt}[1]{\frac{D #1}{Dt}}
\newcommand{\beq}{\begin{equation}}
\newcommand{\eeq}{\end{equation}}
\newcommand{\beqa}{\begin{eqnarray}}
\newcommand{\eeqa}{\end{eqnarray}}
\newcommand{\Grad}[1]{{\bf \nabla} #1}
\newcommand{\Lapl}[1]{{\nabla}^2 #1}
\newcommand{\Div}[1]{{\bf \nabla . #1}}
\newcommand{\Curl}[1]{{\bf \nabla \times #1}}
\newcommand{\B}[1]{{\bf #1}}
\newcommand{\E}{\varepsilon}

\lefthead{Arber, Longbottom \& Van der Linden}
\righthead{Unstable coronal loops}

\begin{document}

   \title{Unstable coronal loops : numerical simulations with predicted
	observational signatures.}

   \author{T. D. Arber, A. W. Longbottom}

   \affil{Department of Mathematical and Computational Sciences,
	University of St. Andrews, KY16 9SS, Scotland. }

   \author{R. A. M. Van der Linden 
		\altaffilmark{1}}
		\altaffiltext{1}{Postdoctoral Research Fellow of the Flemish
			Fund for Scientific Research}

   \affil{Centre for Plasma Astrophysics, KULeuven, Celestijnenlaan 200B,
	3001 Heverlee, Belgium }

\begin{abstract}
	We present numerical studies of the nonlinear, resistive 
	magnetohydrodynamic (MHD) evolution
	of coronal loops. For these simulations we assume that the loops 
	carry no net current,
	as might be expected if the loop had evolved due to
	vortex flows. Furthermore the initial equilibrium is taken to be 
	a cylindrical flux tube with line-tied ends. For a given amount of
	twist in the magnetic field it is well known
	that once such a  loop exceeds a critical length it becomes unstable
	to ideal MHD instabilities. The early evolution of these
	instabilities generates large current concentrations. Firstly we show that
	these current concentrations are consistent with the formation of a
	current sheet. Magnetic reconnection can only occur in the vicinity
	of these current concentrations and we therefore couple the resistivity
	to the local current density.
	This has the advantage of 
	avoiding resistive diffusion in regions where it should be negligible.
	We demonstrate the importance of this procedure by comparison with
	simulations based on a uniform resistivity.
	From our numerical experiments we are able to estimate some 
	observational signatures for unstable coronal loops. 
	These signatures include: the timescale of the loop brightening;
	the temperature increase; the energy released and the predicted observable flow
	speeds. Finally we discuss to what extent these observational signatures
	are consistent with the properties of transient brightening loops.
\end{abstract}


\section{Introduction}
	Coronal magnetic loops have been the subject of considerable observational
	and theoretical study. Of particular relevance here are observations of
	transient brightening loops (\cite{shimizu}).
	Theoretical studies of the stability
	of such loops have hoped to explain some of their observational
	characteristics. In the linear regime the stabilizing effect of line-tied
	boundary conditions has been clearly demonstrated
        (e.g. \cite{hood}; \cite{velli1}). These boundary conditions
	model the fact that on the timescale of ideal MHD instabilities the
	ends of coronal loops can be considered to be frozen into a high-density,
	stationary photosphere. Line-tying allows coronal loops to have a
	more twisted magnetic field, with consequently  more free magnetic
	energy, before the onset of instabilities. However, once a critical amount
	of twist is introduced into the loop it does become unstable
	to ideal MHD modes. 
	The non-linear evolution of these instabilities has 
	been the subject of a number of recent papers. The ideal evolution
	has been studied for a variety of equilibrium profiles
	(\cite{baty1}; \cite{baty2}; \cite{baty3}). 
	The full resistive evolution has also been studied
	(\cite{lionello}; \cite{velli2}) but
	both of these papers assume uniform resistivity. In this paper we extend 
	the work of these papers by performing detailed non-linear simulations
	from both Eulerian and Lagrangian codes.

	All theoretical studies to date have assumed that the initial equilibria
	are one dimensional, i.e. cylinders with photospheric line-tying at each
	end.  We also
	make this initial simplification.
	This has the advantage of greatly simplifying the
	initial equilibrium and allows comparison with previous work. 
	We further restrict our attention to
	loops which carry no net current. This class of equilibrium would result
	from slowly applying a coherent vortex flow to the flux tube.
	This has been shown to lead to a coronal loop of the
	desired form, with a stable input of energy, in numerical experiments
	(\cite{mikic}; \cite{vanhoven}). Such
	an equilibrium, along with the proposed mechanism for its formation, are
	of course greatly simplified, idealisations of the dynamism expected
	in active regions of the solar
	corona. Here many competing effects would be acting simultaneously on each
	loop. Observations show that there is a broad distribution of energy release
	events (\cite{shimizu})
	which are present in active regions. The results here are
	idealised numerical experiments which would be relevant to the large scale,
	high energy tail of this distribution. This paper is more 
	relevant to micro-flares than nano-flares. However, as the distribution
	of flarings/brightenings is very broad it should be understood that for the
	remainder of this paper when we refer to brightening loops we are referring
	to the high energy tail of such events, i.e. compact loop flares or
	micro-flares.

	Several papers have investigated the non-linear evolution of a
	variety of cylindrical equilibria. It is important that the relationship
	between these previous studies and our work is made clear. As this paper
	only considers equilibria which carry no net current it should be 
	understood that when discussing results from these papers we are
	only referring to those results which are for similar equilibria. Two 
	papers (\cite{baty1}; \cite{baty2}) have started
	with unstable coronal loops and run codes with large implicit steps,
	steadily increasing viscosity until all plasma motion is damped. 
	Eventually this process reaches a 
	bifurcated equilibrium state. They have found that in this final
	state the maximum current density generated, $j_{max}$,
	scales with the length of the
	loop. In this paper we resolve the full time-dependent solution and find
	no evidence that $j_{max}$ scales with the loop length. We find that  
	for simulations  without resistivity $j_{max}$ is always the
	largest current possible for the given resolution. This is determined
	by the numerical mesh size in the region of  $j_{max}$, not the loop 
	length, and is true for both the Lagrangian and Eulerian simulations. 
	This does not contradict the results about {\em equilibrium}
	current densities as the $j_{max}$ we observe is not part of an equilibrium.
	Indeed, we observe these large current concentrations at a time when
	there are velocities (resulting from the instability) of the order of the
	local Alfv\'en speed. Such features would have been damped by the numerical
	procedure adopted in searching for equilibrium solutions. The code used
	to find these equilibrium solutions has also been run with a fixed
	viscosity (\cite{baty3}). This paper found that the ideal instability,
	and consequently $j_{max}$, saturated before the current density reached
	the grid scale length. This is in contradiction with the results presented
	here in which there is no evidence for ideal MHD saturation of the
	instability. Furthermore, the resistive phase which is triggered by
	the large current densities in our simulations also prevents the loop
	from ever reaching the ideal MHD bifurcated equilibria described 
	earlier. In this regard we are in agreement with other non-linear
	simulations (\cite{lionello}; \cite{velli2}) which show the same collapse 
	to grid
	scale lengths resulting from the ideal MHD instability, i.e. no signs
	of saturations of the ideal mode. We point out here that while it is
	often assumed that current sheets form as a result of ideal MHD
	instabilities this has never actually been proven. Some papers
	(\cite{baty3}) actually contradict this belief while the only two
	papers which seem to lend weight to this argument 
	(\cite{lionello}; \cite{velli2}) have 
 	only published results for a single resolution. At this resolution
	the current density in the current concentration is only a few
 	times that present in the initial equilibrium. Bearing in mind the lack
	of consensus on this issue we deal with the formation of current sheets
	in some detail in this paper before treating the resistive evolution.
 	The use of a Lagrangian code in this paper allows the formation of current
	densities which are two orders of magnitude larger than has been possible
	in previous studies. We also present detailed tests of the
	scaling of the current density with grid size for both Lagrangian and
	Eulerian simulation. These tests combine to give the most convincing
	evidence to date that current sheets do indeed form as a result
	of ideal MHD instabilities in the corona. In this paper what is meant by
	a current sheet is that unless some dissipative processes is introduced
	into the system the current density will continue to collapse to shorter
        scale lengths without limit. In reality it is resistive effects which 
	stop this collapse.
	It is the treatment of resistivity,
	the greater number of scalings on different grids and tests with
	a Lagrangian code which
	distinguishes our work from previous publications. It is also these 
	differences
	which allows us to make the first quantitative predictions about 
	unstable coronal loops. 
	 	
	While the papers discussed in the previous paragraph have treated a
	variety of equilibria here we study only equilibria which carry
	no net current. We concentrate on a single
	representative equilibrium and perform a variety of
	detailed numerical experiments. In these we abandon the common practice
	of assuming a uniform resistivity and instead couple the resistivity
	to the current. 
	In this model the resistivity is only present in regions where the current
	density exceeds some critical value. In this way resistivity is only applied
	in those regions where reconnection is allowed. 
	This procedure allows us to get closer to a converged answer for real coronal
	values and gives us increased confidence
	in the quantitative predictions of observational signatures which we
	make from our simulations. The evolution
	is largely independent of the form of resistivity used provided
	that it is localised in the current sheet. This point has been postulated
	in other papers (\cite{velli2}) but is confirmed for the
	first time in  our simulations.
	We also include a comparison with 
	simulations in which the resistivity is assumed constant and highlight
	the differences between the two cases. 

	The paper is organised as follows. In \S 2 we describe the physical model
	and explain our motivation for the choice of resistivity. \S 3 contains the
	details of the equilibrium we have chosen along with its linear stability
	properties. The formation of current sheets has been investigated numerically
	and the results of this study are included in \S 4. 
	The full non-linear resistive
	evolution of the loop is outlined in \S 5. Finally the conclusions which we
	can draw from our studies, including predictions of observational 
	signatures of unstable loops, are presented in \S 6. 

\section{Physical Model}
	In this paper we represent coronal loops as initially cylindrical 
	tubes. These tubes have each end tied into the photosphere. 
	Photospheric line-tying is modelled by imposing zero
	velocity at the ends of the loop along with suitable symmetry properties on the
	components of the magnetic field. We ensure that the tube, of length
	$L_z$, is sufficiently long that it is unstable to ideal MHD modes.
	The solution domain is Cartesian with
	transverse size ($L_x$, $L_y$). The instability remains localised
	inside a specific radius so that provided ($L_x$, $L_y$) are large
	enough the choice of boundary conditions
	in $(x,y)$ makes no difference to the evolution of the loop. 

	The evolution of the coronal loop is modelled by the resistive MHD
	equations. In dimensionless form these are
\beqa
\label{continuity}
\dt{\rho}&=&-\Div{(\rho \B{v})} \\
\dt{(\rho \B{v})}&=&-\Div{(\rho \B{v} \B{v})}+(\Curl{\B{B}})\times\B{B}-\Grad{P} 
\label{force}\\
\dt{\B{B}}&=&-\Curl{(\B{v} \times \B{B})} - \Curl{(\eta \Curl{\B{B}})}\\
\dt{\E}&=&-\Div{(\B{v} \E)} - P\, \Div{\B{v}} + \eta \,j^2 \label{energy}
\eeqa
	Where $\B{j}=\Curl{\B{B}}$ is the current density, $\B{v}$ is the velocity
	, $P$ is the thermal pressure, $\E=P/(\Gamma-1)$ is the internal energy
	density ($\Gamma=5/3$ is the specific heat ratio), $\rho$ is the mass
	density and $\eta$ is the resistivity. For all simulations
	we assume a plasma $\beta$ of 0.01. This model
	ignores thermal conduction, radiation, gravity and heating terms other
	than Ohmic. Ignoring the transport terms is justified because of the
	short timescales involved in these simulations. The validity of the
	neglect of gravity is
	of course dependent on the actual length of the loop being considered.
	With pressure scale heights of $\sim 100\,Mm$ in the corona this is
	a good approximation for transient brightening loops but clearly less 
	valid for large quiescent loops. 

	The choice of the functional form for the resistivity $\eta$ in the
	simulations presented in this paper is particularly important. 
	Normally $\eta$ is treated as a constant
	chosen on numerical grounds such that the explicitly included resistive
	diffusion exceeds numerical diffusion. In this paper $\eta$ is chosen
        so that it is only included in those regions where it is needed. The
	classical resistivity of coronal plasmas is negligible on the 
	timescale of MHD instabilities, i.e. the timescale of interest here,
	everywhere except in regions of intense current concentration. Hence
	we choose a resistivity given by
\beq
\eta=\eta_0 \, MAX(0, |j| - j_{crit})
\label{eta}
\eeq
	In this formula $\eta=0$ if $|j| \le j_{crit}$ where $j_{crit}$
	is the critical value of current density needed before the resistivity
	is turned on. Thus while $\eta_0$ is still chosen on numerical grounds
	the resistive effects are only applied to those regions in which they
	are actually needed. The point 
	here is essentially that if we were to take a uniform resistivity we would
	be diffusing the magnetic field in regions were resistivity should be
	negligible. It is true that {\em locally} the effect of uniform
	resistivity would be largest in the current sheet and in that region 
        the difference due to the
	choice of uniform resistivity over equation \ref{eta} would be small. However,
	the small resistive effects which are then applied over a much larger total
	volume do change the nature of the final solution. This will be discussed
	in \S 5 where a comparison of the two approaches is presented. 

	If intense current concentrations form as a result of the ideal MHD
	instability then the electron fluid flow speed $U_e$ will become large.
	These current concentrations lead to localised heating of the electrons
	and on the timescales of interest here these are not in thermal equilibrium
	with the ions. Thus the electron temperature will exceed the ion's 
	and once $U_e$ exceeds the local ion sound speed the current concentration
	will drive ion-acoustic turbulence. The fluctuating electric field of
	this turbulence causes electron scattering which is manifested at the
	fluid level as an increase in the local resistivity (\cite{rosner}).
	By rapidly creating a current sheet the ideal MHD instability therefore
	creates all of the conditions necessary for the onset of ion-acoustic
	turbulence. These are that the electron temperature exceeds the ion
        temperature and that the electron flow speed exceeds the local ion
        sound speed. While these conditions are satisfied for the current sheets
        driven by ideal MHD instabilities it must be noted that in other 
        circumstances in which current sheets form this may not be the case.
	If the electron flow speed continues to be driven, as is the
	case here, this enhanced resistivity increases. Equation \ref{eta}
	would then be a suitable macroscopic parameterisation of the sub-grid
	scale turbulence.
	For a coronal loop with number density 
	$10^{16}\,m^{-3}$; ion thermal speed $1.3 \times 10^5\,ms^{-1}$;
	width $10^6 \,m$ and magnetic field strength $100 \,G$
	the normalised critical current density required for the onset
	of such turbulence is $j_{crit} \sim 3000$.
	These are order of magnitude estimates of coronal values appropriate 
	for transient brightening loops (\cite{shimizu2}).
	Unless otherwise stated these characteristic values for density etc.
	are the ones used throughout the paper when estimating real, i.e.
	unnormalised, values.  Note
	however that the temperature assumed in calculating the thermal speed
	is $2\times 10^6\,K$. This is taken as an average temperature of an
	active region and is meant as an estimate of the loop temperature
	{\em before} the brightening occurs. Of course the codes use normalised
	variables so scaling to other coronal values is a trivial matter
	and does not affect the qualitative features of these simulations. 
	This estimate of $j_{crit}$ should be
	compared with the equilibrium current density which has a maximum value
	of 4.5 in these units. When ion acoustic turbulence is active the
	characteristic electron scattering time is approximately the ion plasma
	period. This gives an increase in the plasma resistivity by a factor
	of up to $10^6$. If one uses the Spitzer formula for classical
	resistivity one concludes that the normalised resistivity is
	$\sim 10^{-12}$. However in current sheets which trigger ion-acoustic
	turbulence $\eta_0$ in equation (5) is $\sim 10^{-6}$.
	It should also be
	pointed out that ion-acoustic turbulence is not the only possible
	source of enhanced resistivity. For example the initial rapid heating
	in the current sheets of unstable loops has been shown to trigger
	Langmuir wave enhanced resistivity (\cite{takakura}). While there is
	theoretical evidence that the mechanism described above would lead
 	to enhanced resistivity there is no direct evidence that such levels
	of turbulence are indeed generated and sustained on the required scale 
	in current sheets in the solar corona. As a result the arguments
	in support of the use of Equation \ref{eta} as the correct {\em physical}
	form of resistivity, as opposed to being simply {\em numerically} 
	appropriate as discussed in the previous paragraph, must remain  
	speculative at this stage.

\section{Equilibrium and Linear Stability}

	Throughout this paper we limit our attention to just one equilibrium.
	This has allowed us to perform a detailed set of numerical experiments,
	over a range of grid sizes, using different codes. The equilibrium is
	taken to be a force-free cylinder which carries no net current. The
	precise form is given in terms of the axial current density in the loop,
	$j_z$, by
\beq
j_z = j_0\left(1-\frac{r^2}{b^2}+a\frac{r^3}{b^3}\right)
\eeq
	The normalisation used is that the loop is confined within a radius $r=1$
	so that the free parameters $a$ and $b$ are found from the conditions
	$j_z(1)=0$ and $\int_0^1 rj_z(r)\,dr=0$. These guarantee that 
	the equilibrium
	does not contain a surface current at $r=1$ and that the total current in the
	loop is zero. The actual values are $a=5/(6\sqrt{6})$ and $b=1/\sqrt{6}$.
	The poloidal component of the magnetic field,
	$B_{\theta}$, is then found from Ampere's law and $B_z$ from the
	force-free condition,
\beq
B_z^2=B_{0}^2 -  B_{\theta}^2 - 2\int_0^r\left(\frac{B_{\theta}^2}{r'}\right)\,dr'
\eeq
	This equilibrium definition has two free parameters, $B_0$ and $j_0$.
	These are chosen to be 1.0 and 4.3 respectively. With this choice the 
	equilibrium profiles are shown in Figure \ref{fig1}. Note that outside 
	$r=1$
	the magnetic field is purely axial. 
	This choice of equilibrium
	is similar to  equilibria with no net current studied elsewhere (\cite{velli2}; 
	\cite{baty3}; \cite{lionello}). These works all showed that once the
	critical length for the onset of instability has been exceeded the growth
	rate increases rapidly, eventually reaching a growth rate which is almost 
	independent of length. 
	Our choice of
	$L_z=10$ for most of our simulations puts the equilibrium in this region. 
	We therefore do not self-consistently
	follow the loop as, for example, its length increases from a stable length
	up to the length considered here.
	We simply assume that it is initialised
	at a length which is already unstable at the
	saturated growth rate. This is necessary 
	numerically so that the instability grows sufficiently quickly to be simulated
	in a reasonable time. We have performed some simulations for shorter 
	initial lengths
	and found that our predicted observational signatures are insensitive to this
	choice. 
	As far as the ideal phase of our
	simulations are concerned we show broad agreement with the results in 
	earlier works. This shows that these results are not critically sensitive
	to the details of the equilibrium and is a useful confirmation of the 
	results
	from a completely different set of codes. This paper differs from those
	earlier publications in performing more detailed scaling tests on the
	formation of the current sheet and more detailed simulations of the 
	resistive phase. The choice of equation \ref{eta} for the resistivity
	is a particularly important difference between this work and previous
	publications.

	In the first instance we studied the linear ideal MHD stability of our model
	equilibrium. The MHD equations (\ref{continuity})-(\ref{energy}) for
	zero resistivity are linearised, and a time dependence 
	$\propto \exp (\gamma t)$ assumed
	to obtain the equations for the linear normal mode spectrum in ideal 
	MHD. These normal modes are then calculated using a bicubic finite
	element code similar to that described in \cite{vh98}. The qualitative
	features of the stability of this equilibrium are the same as those for
	previously studied equilibrium profiles, e.g. \cite{lionello}.
	Unless otherwise stated the length of the loop in the remainder of
        this paper will be
	fixed at $L_{z}=10$. This point is chosen some distance from the
	marginal stability point, so that the growth rate of the fundamental
	mode ($\gamma\approx 0.302$) is nearly equal to its 
	maximal value. This allows the instability to develop sufficiently
	rapidly to save computational overhead. At this loop length, two more
	instabilities exist: the first overtone has a growth rate of $0.197$,
	while the second has growth rate $0.037$. The latter is certainly not
	expected to have any significant contribution to the development of
	the instability due to its small growth rate, while the former might
	contribute, but in practice does not
	show up in the time-dependent simulations presented in this paper
	(which is mainly due to the choice of initial perturbation).
	
	To make a clear distinction between the growth of the linear instability
	and the effects of non-linearity on its evolution, it is very useful to
	compare both linear and non-linear time-evolution of the same initial
	perturbation. A good approximation of the linear time-evolution may be
	obtained from the spectrum of normal modes by writing the initial
	perturbation as a linear combination of the full spectrum of normal
	modes.
	A quantitative comparison of the linear versus non-linear evolution
	is made at $t=5$ in Figure
	\ref{fig2}, where the perturbed current density is plotted along  
	a line through the centre of the loop parallel to the photospheric line-tied
	ends
	from the linear and both non-linear simulations. It is clear that at this
	time, the evolution of the instability is still linear in nature.
	In Figure \ref{fig3} a similar comparison is made at the later time
	of $t=10$. 
	At the early time, i.e. $t=5$, the linear and non-linear results differ
	in two noticeable ways. Firstly the peaks from the non-linear analysis
	are shifted to the right due to the central plasma column moving in the
	$x$ direction. 
	Figure \ref{fig4} shows $v_x$ along the same $x$-axis at $t=5$ from the
	linear code. From this one can see that the regions with the steepest
	gradients are located in the region $0.6\le |x|\le 0.8$. These are
	regions of compression for $x>0$ and expansion for $x<0$. They
	account for the asymmetry seen in both non-linear results. By $t=10$ the
	general structure of the linear mode is still evident but non-linear
	effects are becoming significant. Most prominent is the formation of a
	large current density at $x=0.8$. Note that Figure \ref{fig3} is truncated
	at $|j_1|=10$ and that for the Lagrangian code $|j_1|^{max}=31$ at $t=10$
	
\section{The Formation of Current Sheets}

        The fully non-linear evolution of the instability is followed using two
        time-dependent nonlinear MHD codes. The first is an ideal MHD Lagrangian code 
        that follows the initial phase and the formation of the current sheet, the 
        second a resistive MHD Eulerian code that allows the resulting reconnection 
        to be followed and the later phases of the evolution examined.

        The Lagrangian code is based on the equilibrium code of \cite{longbottom}
        and is described in the appendix. It has two main features relevant to this 
        study. The code solves the ideal MHD equations (there is no dissipation due 
        to resistivity or viscosity). The grid on which the equations are solved 
        moves with the fluid and thus in this case, where the inner region of the
        loop is forced out against the near stationary outer potential field, more
        grid points will accumulate at the regions where large gradients form. These 
        two features together ensure that the current structures resulting from 
        the instability remain highly resolved. 

        The loop is confined inside $r=1$ and centred in the $(x,y)$ computational
        plane which is $L_x=6$ by $L_y=6$. The simulations were carried out on
        grids of $61 \times 61 \times 21$, $91 \times 91 \times 31$ and
        $151 \times 151 \times 101$ points in the $(x,y,z)$ directions. The $(x,y)$ 
        grid is uniform inside $-1.1\le x,y \le 1.1$ and only stretched outside 
	this central core,
        with the initial grid spacing (at $t=0$) within $-1.1\le x,y \le 1.1$
	being 0.05, 0.033 
        and 0.02 respectively. The $z$ grid is uniform with $-5 < z < 5$. Simulations 
        are started with a small velocity perturbation ($v_{max}=0.01$) whose
        structure is taken to approximate that found from the linear analysis. The
        results below are shown for $151 \times 151 \times 101$ grid points.

        As described in the previous section the initial evolution agrees well
        with that predicted by linear theory, the linear eigenfunction and 
        growth rate being reproduced. However, from $t=5$ onwards a 
        helical current structure grows as a result of the inner twisted magnetic 
        field being forced against the surrounding potential field by the developing
        kink mode. This behaviour can be seen in Figure \ref{fig5} which shows the 
        perturbed current ($|j_1|$) plotted along the x-axis at $y=z=0$ as a function 
        of time. Here the maximum value of current plotted has been truncated at 
        $j_1=50$ so that the details at earlier times can be seen. The actual maximum 
        value of current in the current sheet at $t=11.5$ is $j_1=766$. Even at late
        times the remnant of the linear mode are still visible in the central part of 
        the loop. The formation of the current structure at the rational surface can be
        clearly seen. The maximum current after $t=10$ scales faster than $n_x^2$,
        where $n_x$ is the number of gridpoints in the $x$ direction in the 
	Lagrangian code. 
        No sign of saturation of the current is seen. This behaviour is indicative of 
        the formation of current sheets (\cite{longbottom}).

        A surface plot of the total current in the $(x,y)$ plane at the loop apex is 
        shown in Figure \ref{fig6}. Again the current has been truncated at 
        $j=50$ so 
        that both the shape of the current concentration and the inner structure is 
        visible. The global structure is that of a helix wrapped around the central 
        loop column and is essentially identical to that found by the Eulerian code,
        Figure \ref{fig7}a. The half-width of the current sheet at $t=11.5$ is 
        approximately 0.0025. This would mean that for the Eulerian code to 
	resolve this 
        current at this time there would need to be $\sim 3500$ grid points 
	in both the $x$ and the $y$ directions.

	These results have also been confirmed with the Eulerian code described
	in detail in the next section. The important point as far as
	the current sheet formation is concerned is that  
	running in the ideal MHD model, i.e. with $\eta=0$, the
	current generated in the current sheet scaled as $1/dx$,
	where $dx$ is the grid resolution in the current sheet, and showed no
	signs of saturating as the resolution was increased. Note that the scaling
	of the maximum current is different than in the Lagrangian code as the
	Lagrangian code moves points into the region where the current sheet is
	formed. For the Eulerian code the
	highest current obtained in the current sheet was $\sim 30$. 
	This behaviour is consistent with previous work in this area
	(\cite{lionello}; \cite{velli2}) although by performing scalings
	on different grid resolutions we are able to verify the correct scaling
	with mesh size which is required if these are indeed current sheets.
	Furthermore the current densities generated in the Lagrangian code
	are two orders of magnitude larger than has been possible with the
	Eulerian codes used in previous studies. 
	The results in this paper contradict some simulations which
	found non-linear saturation of the instability with a purely ideal MHD
	description (\cite{baty3}). This discrepancy between different codes has
	been noted before (\cite{baty3}; \cite{lionello}) and has been attributed
	to the different treatment of small scales, with associated numerical
	dissipation, in the codes used. Here we have used two non-linear codes,
	each with distinctively different properties from earlier codes, and
	found no saturation. This is the first time a Lagrangian code has been
	for such simulations and that tests on the scaling of the current
	density with $dx$ have been presented. These provide strong 
	evidence that these instabilities do not saturate while still described
	by ideal MHD and that the generated current concentrations are current
	sheets.
	We have also performed simulations of loops with lengths $L_z=5.5$ 
	finding the same magnitude current sheet evolving in both cases.
	Previous studies of bifurcated equilibria (\cite{baty1}; \cite{baty2})
	have found that $j_{max}$ scales linearly with $L_z$. Our simulations
	show that this is not true of the instability driven current sheet 
	which forms while the loop is still in a highly dynamic, non-equilibrium
	state. 

\section{The Resistive Phase}
 
	The Lagrangian code is only valid for ideal MHD. To follow the resistive
	evolution of the loop an Eulerian code is used. The ideal MHD part of
	this code is based on the MH3D code (\cite{lucek}) written
	at Imperial College, London. The code uses a stretched Eulerian grid.
	Variables on this fixed grid are updated using a split Lagrangian, Eulerian
	remap technique, with the advection handled by a
	second order Van Leer upwind scheme (\cite{youngs}). The code maintains
	$\Div{\B{B}=0}$ by using the constrained transport model for magnetic
	flux advection (\cite{evans}). During the Lagrangian phase of each
	timestep an artificial viscosity term is added to equation \ref{force}.
	This is applied as a viscous pressure at cell boundaries for
	cells which are being compressed (\cite{richtmyer}). This viscosity results
	purely from compressive effects, i.e. shear viscosity is not included.
	For comparison
	with papers which add a viscous term of the form $\nu \rho \Lapl{\B{v}}$
	to equation \ref{force} the formula we use is approximately equivalent to
	this form with $\nu=10^{-3}$. We have confirmed this with direct comparison
	of the two forms on low resolution runs. It should also be
	pointed out that in our units $\nu=10^{-3}$ is the correct order
	of magnitude for a transient brightening loop. The viscous heating in these
	simulations is significant in the overall energy balance and so must be
	included in the energy equation.

	The results presented here are from runs with a $161^3$ Cartesian grid.
	This is the resolution used in the largest set of numerical experiments
	and therefore constitutes our largest consistent data set.
	Some higher resolution tests have been performed to test convergence.
	These had a $221\times 221\times 101$ grid in $(x,y,z)$ with the grid
	stretched to give twice the resolution in $(x,y)$ of the $161^3$
	experiments. Results from this resolution will be called the high 
	resolution results in the remainder of this paper.
	Unless explicitly stated it should be assumed that results
	are from the $161^3$ runs.
	The loop is confined inside $r=1$ and centred in the $(x,y)$ computational
	plane which is $L_x=6$ by $L_y=6$. The $(x,y)$ grid is uniform inside
	$-1.1 < x,y < 1.1$ and only stretched outside this central core. The
	ratio of minimum to maximum grid spacing was 3.2. The $z$ grid was
	uniform with the coordinate range $-5 < z < 5$. Simulations are
	started with a small velocity perturbation ($v_{max}=0.01$) whose 
	structure is taken to approximate that found from the linear analysis. 
	
	Figure \ref{fig7} shows iso-surfaces of the magnitude of the current
	density at two different times. The surfaces are iso-surfaces of $|j|=3$.
	As the maximum current in the initial equilibrium is 4.5 these surfaces
	show the perturbed central column as well as the current sheet. While
	current density is not experimentally observable these figures are presented
	as they give the clearest picture of the physical processes present in
	this unstable loop. These results are from a simulation with $\eta_0=10^{-3}$
	and $j_{crit}=5$. The value of $\eta_0$ is the smallest value we can use
	in this code and still guarantee that the deliberately included, numerically
	controlled resistivity is larger than the numerical resistivity inherent in
	the difference scheme. $j_{crit}$ is fixed to the largest value that allows
	us to fully resolve the current sheet. The non-linear feedback through
	resistivity of the form given in equation \ref{eta} then restricts the
	maximum current in our current sheets to about 10. Figure \ref{fig7}(a)
	shows the iso-surface at $t=10$ with the central column perturbed into the
	characteristic helical $m=1$ mode. Wrapped around this central
	column is the current sheet which is formed at the place where the pitch
	of the instability matches the pitch of the magnetic field (see \cite{baty1}
	for a detailed discussion of this process). At this time the current in this
	outer current sheet has just reached 5. Therefore up to this point equation
	\ref{eta} has been setting $\eta=0$ everywhere 
	and the code has been solving the ideal
	MHD equations. Beyond $t=10$ the current in the current sheet continues to 
	increase and equation \ref{eta} `turns on' the resistivity and the code 
	automatically begins solving the resistive MHD equations but with the 
	resistivity
	only present in the outer current sheet. Figure \ref{fig7}(b) shows
	the current iso-surface at $t=15$. More of the central column has been 
	moved out, by the ideal MHD instability, into the region of the current
	sheet. In this region reconnection is allowed so that the twist in field
	lines can be removed, or equivalently the current dissipated. This process
	continues until at $t=20$ sufficient current has been dissipated that
	no region has $|j|>5$ and the resistivity
	'turns off'. In summary Figure \ref{fig7} shows that the ideal MHD
	instability drives the current
	in the loop out into the current sheet were it is dissipated. 

	The above experiment has been run with two different plasma 
	density profiles. In both
	cases the pressure is uniform, as required by the force-free condition.
	In one set of tests the plasma density was taken to be uniform across the whole
	computational domain. In the other the density profile was taken to be
	$\rho=0.45(1+\cos (\pi r))+0.1$ for $r\le 1$ and $\rho=0.1$ for $r>1$. This
	second choices makes the average density inside the loop 3.67 times that
	of the surrounding coronal plasma and was motivated by observations that
	the density inside a brightening loop exceeds that of the surrounding coronal
	plasma. 
	These experiments showed that the evolution
	of the unstable loop is insensitive to the choice of density profile. It should
	be noted that while the second density profile does imply a drop in temperature
	inside the loop this is unimportant for MHD simulations. It is the 
	pressure which exerts a force and the temperature does not appear in equations
	(1)--(4). 
	Changing the density merely changes the timescales
	involved. What is clear from the second density profile is that in the final 
	state, i.e. at $t=20$, the density enhanced region still lies within the same
	bounding radius. In other words the loop is not destroyed by the instability
	but the twisted  magnetic field lines are straightened out. This is in agreement
	with previous studies (\cite{lionello}).

	Figure \ref{fig8} shows two different iso-surfaces of energy density
	taken at the same time, $t=20$. Figure \ref{fig8}(a) shows regions which
	have been heated to 3 times the initial background value. This shows that
	the loop has brightened along its whole length. Figure \ref{fig8}(b) shows
	the higher energy components  which have been heated to 6
	times the background value. In this Figure the central region has been heated
	by Ohmic dissipation in the current sheet while the ends have been heated by
	the viscosity in the code. The rings at each end of Figure \ref{fig8}(b) 
	are where the loop is tied into the photosphere. The very dynamic nature of
	the loop at $t=20$, see below, causes viscous stresses at the ends where the
	velocity is forced to be zero by the photospheric boundary conditions.	
	Once the current density drops below $j_{crit}$ everywhere the resistive
	phase is over and the code reverts to ideal MHD (although viscosity 
	is still present). 
	No simulations have been
	performed beyond this time as from that time onwards the timescale of
	interest is the timescale for thermal conduction. This
	is too long to be studied in 3D with this kind of resolution. 

	At $t=20$
	there are very large flows set up due to the ideal instability and 
	field line reconnection. The peak value is $\sim 0.8 \,V_A$, where $V_A$ is
	the Alfv\'en speed. However, observations of flows in the corona are based
	on Doppler shift measurements. Such measurements include averaging over
	exposure times, pixel sizes and line of sight effects. To estimate
	the importance of these effects we have taken a simple
	density weighted average of one of the transverse components of velocity,
	i.e. $v_x$ or $v_y$, over an area of approximately $1.5\,Mm \times 1.5\,Mm$,
	an exposure time of 10 seconds and along a line of sight.
	This averaging time and area are typical of solar observations. The flow
	structure after this averaging is shown in Figure \ref{fig9}. In this figure
	the photospheric footpoints are at $\pm 5 Mm$ and the points are the centres
	of our pixels. 
	This sort of simple averaging can only be taken as an estimate
	of the kind of velocities which could be observable experimentally. Issues
	such as ionisation population levels, temperature dependence of the weighting
	function are beyond the scope of the current work. Figure \ref{fig9} does 
	however show that the very large flow speeds present in our simulations
	would not be directly observable. 
	These simulations therefore suggest that
	after a loop with parameters typical of a brightening loop has gone unstable,
	flows of $\sim 40 km\,s^{-1}$ should be observable. The averaging implicit in
	these measurements would however mask the real loop plasma flows which are
	highly localised and as large as $1500 km\,s^{-1}$. 
	  
	The energy released from this instability is 
	54\%  of the available magnetic energy.
	This is split almost equally between Ohmic heating, kinetic energy
	and viscous heating. 	
	The available energy is defined as the energy
	stored in the $B_{\theta}$ component of the equilibrium magnetic field.
	For this equilibrium the free magnetic energy for the coronal values
	in \S 2 is $9.5\times10^{28}\,ergs$. 
	For typical brightening loop values the resistive phase lasts about 
	5 seconds.

	For comparison we have repeated the above simulation with a uniform 
	resistivity. This is turned on at $t=10$ and the simulation is stopped
	at $t=20$. In this way the resistivity is applied for the same length of
	time as above. For this run we took $\eta=10^{-3}$ so that this too is
	consistent with the value used above. We find that for this uniform
	resistivity model 62\% of the available energy is released. 
	This is sufficiently close to the value found from using equation \ref{eta}
	that the difference can be ignored. However, the kinetic energy generated with
	a uniform resistivity is approximately half
	that of the value from using equation \ref{eta} and the
	total Ohmic heating is three times larger. The peak flows are
	less than half those shown in Figure \ref{fig9}. At present all
	simulations are forced to use resistivity which is unphysically large.
	These tests show that applying such a large resistivity {\em uniformly} over
	the computational domain, instead of localising it to just those regions
	where it should have an effect, over-estimates the Ohmic heating
	and under-estimates the kinetic energy in the final dynamic state. 
	Neither of these points is surprising as including resistivity everywhere
	will clearly increase the overall Ohmic dissipation and smooth the fields
	driving the instability. What is important here is that these effects
	have now been quantified for the values of resistivity typically used
	in large scale numerical simulations. Using a uniform resistivity does
	release the same amount of magnetic energy
	as using equation \ref{eta} but splits it between Ohmic heating and
	kinetic energy in a very different way. It is worth noting that runs
	with {\em zero} resistivity also release the same amount of energy but
	most of this energy is simply lost from the system. 
	In this case when 
	the current density scale length reaches the grid spacing {\em numerical} 
	diffusion dissipates the current. Thus getting the correct amount of
	total magnetic energy released in a simulation is not of itself a
	useful indicator that the resistive effects have been correctly modelled.

	All of the results above are from $161^3$ stretched grids with 
	$\eta_0=10^{-3}$ and $j_{crit}=5$. By running the same simulations
	on $81^3$ and $121^3$ grids we have confirmed that, for these values
	of $\eta_0$ and  $j_{crit}$, these results are the correct,
	converged solutions. We have also conducted three higher resolution
	simulations on a stretched $221\times 221\times 101$ grid. The first was
	a purely ideal MHD simulation. This confirmed that the current density in
	the current sheet scales as $1/dx$ demonstrating
	that we have no evidence of the current density saturating consistent with
	the results from the Lagrangian code. The second simulation repeated
	previous resistive runs with $\eta_0=10^{-3}$ and $j_{crit}=5$. This confirmed
	the accuracy and convergence of these results. The last high resolution
	run repeated this last simulation but with $j_{crit}=10$. It is only
	at this higher resolution that enough grid points are present in the
	current sheet for energy conservation to be acceptable, i.e. the Ohmic
	heating is much larger than the energy loss through numerical diffusion,
	for this value of $j_{crit}$.
	The interesting point here is that the results with $j_{crit}=10$ are
	not significantly different, i.e. none of the observational signatures
	change, from those with $j_{crit}=5$. Similarly, increasing $\eta_0$ to
	$10^{-2}$ does not alter our observational predictions. Hence over the
	range of dimensionless parameters resolvable by the simulations we
	have performed we find that the predicted observational signatures are
	insensitive to the choice of $\eta_0$ and $j_{crit}$. However, the use
	of equation \ref{eta} for coronal values assuming that turbulence
	enhanced resistivity (as discussed in \S 2) was active would require 100 times
	the resolution we have used. If such enhanced resistivity were absent then
	of course much higher resolution still would be required.
	Such grid sizes are of course impossible
	at present. Our predicted signatures are therefore based on the assumption
	that this independence on $\eta_0$ and $j_{crit}$ remains true up to
	coronal values. 

	The results above are for a loop with $L_z=10$. We have also checked this
	result for $L_z=5.5$ which has approximately half the growth rate and is
	closer to the marginal stability length. The resistive evolution of this
	loop showed the same behaviour as for $L_z=10$. The maximum current
	in the current sheet was the same; the timescale for the resistive phase
	was 10 Alfv\'en transit times as before and the energy released was the
	same fraction of the available free energy. The only difference was that
	as a result of the lower growth rate it took longer to reach the stage where
	the current density in the current sheet triggered the resistivity. A
	final set of tests has also been performed in which the resistivity was
	set to a constant value of $10^{-3}$ in any computational cell with
	current density larger that $j_{crit}$. These also gave the same set
	of observational signatures as presented above verifying that the results
	are also insensitive to the functional form chosen for the resistivity
	provided that it is only present in the current sheet. 

\section{Discussion}

	Our aim in this work has been to perform non-linear numerical solutions
	of unstable coronal loops using resistive MHD. In these simulations we
	have concentrated on a single equilibrium. We have confirmed some of the
	results of previous papers using different initial conditions and different
	numerical approaches. Where there has been a lack of consensus in
	previous studies we have clarified these issues by supplying detailed
	numerical results.
	These include the formation of current sheets; the
	basic structure of those current sheets and the general features of the
	resistive evolution. There are however important quantitative differences
	between this paper and previous work. Most of these differences stem from 
	only applying resistivity locally in current sheets. Before progressing to
	a description of the observational signatures it is worth
	highlighting what those differences are.
\begin{enumerate}
	\item We have run the non-linear codes on a range of grids. These confirm
	that the current generated near the resonance surface scales as	
	$1/dx$ for the Eulerian code and faster than $1/dx^2$ for the Lagrangian
	code.
	The current densities
	found in the Lagrangian code are two orders of magnitude large than 
	those found in previous studies. This combined with the above scalings
	with grid size provide convincing evidence that the current density
	would continue to collapse to smaller and smaller scales unless stopped
 	by some dissipative process, i.e. they are current sheets.
	\item Coronal simulations require the use of a resistivity which is larger
	than the real coronal value in order to limit the current densities formed
	in such simulations. We adopt this procedure but only apply
	resistivity where it is needed by using equation \ref{eta}
	to localise resistive effects to regions with high current densities.
	If the current density exceeds 3000 (in our normalised units) 
	then equation \ref{eta} may be viewed as a parameterisation of the 
	effects of sub-grid
	scale turbulence  enhanced resistivity based on theories of 
	ion-acoustic turbulence. While there is theoretical evidence that this
	may be true a direct proof of this is beyond the scope of this paper.
	\item We have quantified the discrepancies between using equation \ref{eta}
	and taking $\eta$ to be uniform over the computational domain. 
	\item For all of the tests we have run, the observable properties of the
	loop are independent of the choice of $j_{crit}$ and $\eta_0$ in equation 
	\ref{eta}. As far as we can determine it is the rate at which magnetic flux
	is moved into the reconnection region which is important. This is determined
	by the correct resolution of the ideal MHD instability. We have also shown
	that the results are insensitive to the function form of equation 5.	
\end{enumerate}

	The simulations we have performed are only for one equilibrium. This equilibrium
	carries no net current. Such an equilibrium might evolve due to long scale
	length correlated twisting of the flux tube either in its rise through the
	convection zone or through photospheric vortex flows once it has emerged into
	the corona. We further assume that the equilibrium is force-free and that at
	the start of our simulation the loop is unstable. The loop could have become
	unstable from increased twisting of the magnetic field due to photospheric
	motion or from rising higher into the corona and hence increasing in length.
	Provided these conditions are satisfied, or at least a reasonable 
	approximation, then the observational signature of the ensuing instability
	are those set out below. Where results are presented in unnormalised units
	it is assumed that they relate to the typical brightening loop values listed
	in \S 2. While we have shown that our results are independent of $\eta_0$
	and $j_{crit}$ over a large range of values it should be noted that
	these observational signatures are only valid if we assume that this remains
	true when extrapolating to real coronal values. Implicit in this 
	is of course the assumption that the microscopic details of magnetic field
	diffusion and reconnection can be adequately parameterised on the fluid
	level in terms of a scalar resistivity. We have no reason to doubt this
	but of course a proof is beyond the scope of these, or any currently available,
	numerical simulations. 
\begin{enumerate}
	\item The instability will trigger the formation of an intense current
	concentration (a current sheet). The combination of instability and
	current sheet dissipation
	will cause the loop temperature to increase along
	its whole length by a factor of 3 over the initial background value,
	i.e. heating up to about $6\times 10^6\,K$. A higher
	temperature component, perhaps 6 times the initial temperature
	(around $1.8\times 10^7\,K$), may also be
	observable (see Figure \ref{fig8}(b)).
	\item The whole resistive phase takes about 10 Alfv\'en transit times
	(approximately 5 seconds) so the loop would brighten very rapidly.
	The viscous dissipation timescale for this loop is $\tau_{\nu}\sim 4$ minutes.
	In this we have taken $\tau_{\nu}=L^2_v/\nu$ where $L_v$ is the velocity
	scale length and we have taken $L_v=10^6\,m$. 
	The conductive timescale (the timescale for which the loop should be visible
	at 3 times background temperature) is also of the order of minutes.
	\item For this equilibrium the total magnetic energy released is
	$\sim 5\times 10^{28} \, ergs$. This is split approximately equally
	between Ohmic heating, viscous heating and kinetic energy. The energy released
	for other size loops can be found from noting that the energy released
	scales as $B_0^2 a_0^2 L_z$ where $B_0$ is the magnetic field in the loop,
	$a_0$ is the loop radius and $L_z$ the loop length.
	\item The loop is not destroyed but remains confined within the same region
	of the corona. The twisted field lines become straightened within the same
	confining region.
	\item After brightening there will be very large, localised flows in the loop.
	These have a maximum of $\sim 0.8 V_A$ ($\sim 1500 km\,s^{-1}$). However,
	after taking into account the line of sight effect; exposure times and averaging
	over a diagnostic pixel size we find that the predicted {\em observable} 
  	flows are
	$\sim 40 km\,s^{-1}$. The timescale of viscous dissipation means that these
	flows would persist for minutes after the initial brightening and therefore
	should be observable. 
	\item The kinetic energy generated is about half the total thermal energy 
	released (both Ohmic and viscous heating) and this may
	be indirectly observable. The averaging inherent in measurements outlined
	above would mean that temperature estimates based on spectral line 
	widths (which would included the averaged Doppler broadening) should be
	about 1.5 times those from calculations based on ratios of line intensities
	(which will be insensitive to the plasma motion predicted here).   
\end{enumerate}
	
	The clearest study of transient brightening loops (\cite{shimizu}) shows
	that the above diagnostic signatures are consistent with some of the
	SXT observations from Yohkoh. The energy released by the instability is
	at the high end of observed energies for brightening loops, i.e. they are
	micro-flares. This is as should be expected since the idealised nature of these
	simulations is only applicable to loops which have been twisted 
	by  a long timescale 
	input of energy. The majority of these loops are seen to brighten along
	their entire length consistent with the picture in Figure \ref{fig8}(a).
	Assuming a background active region coronal temperature of $2\times 10^6 \, K$
	the iso-surfaces in Figure \ref{fig8}(a) would correspond to temperatures
	of $\sim 6\times 10^6 \, K$. The more structured, high temperature component
	shown in Figure \ref{fig8}(b) would correspond to a temperature of
	$\sim 2\times 10^7 \, K$.
 	The conductive loss timescale for a coronal
	loop at $\sim 10^6\,K$ is $\sim 10^2 s$, so one would expect the brightening
	of the loops predicted here to last several minutes. However the very high 
	temperature
	component is highly localised (see Figure \ref{fig8}) and would
	be smoothed out on a much more rapid timescale. 
	It would therefore require an exposure time of the order of  
	seconds, on a diagnostic sensitive to temperatures $\sim 2\times 10^7 \, K$ 
	to confirm or dismiss the predictions shown in Figure \ref{fig8}(b).
	The speeds predicted in these simulations are consistent with observed speeds
	from Doppler measurements but a more directed study is needed to confirm
	if the size and structure shown in Figure \ref{fig9} are present
	{\em after} a loop brightening event. The predicted
	discrepancy between temperature measurements based on line broadening
	and ratios of different spectral line intensities is also as yet untested. 

	In summary, we have performed a set of numerical simulations of unstable
	coronal loops which carry no net current. By using high resolution numerical
	experiments, with a resistivity given by equation \ref{eta}, we have been able
	to have greater confidence in our {\em quantitative} predictions from these
	simulations than would have been possible by simply assuming uniform
	resistivity. Our predictions for these observational signatures are listed above. 
	Where comparison with current observational data is possible
	these are in broad agreement with the properties of brightening (compact
	flare) loops. The full set of predictions can now be used as the basis 
	of a directed observational study which may then confirm, or dismiss, whether
	large scale MHD instabilities are the cause of the high energy,
	micro-flare end of the spectrum of transient loop brightenings.

\acknowledgments
	This work was supported in part by a PPARC rolling grant at St. Andrews and 
	by a mobility allowance granted to one of the authors by the Flemish
	Fund for Scientific Research. 
	The authors would also like to thank Alan Hood for many helpful discussions,
	A. R. Bell of Imperial College for making his MH3D code 
	available to us and to S. G. Lucek, also from Imperial College, for
	invaluable discussions about the use of the MH3D code. 

\appendix
\section{Details of the Lagrangian Code}
        The Lagrangian code used to follow the ideal evolution of the instability
        is based on the equilibrium code described in \cite{longbottom} and 
        \cite{craig}. It solves the ideal Lagrangian MHD equations
\beqa
\rho \Dt{\B{v}}&=&(\Curl{\B{B}})\times\B{B}-\Grad{P}, \label{lag_mom}\\
\Dt{\B{x}}&=&\B{v}, \label{lag_pos}
\eeqa
        with 
\beqa
\left(\frac{P}{\rho^{\Gamma}}\right)&=&\hbox{constant (moving with the fluid)},
     \label{lag_eng}\\
\rho&=&\rho_0/\Delta, \label{lag_rho}\\
B_i &=&\frac{\partial x_i}{\partial X_j} B_0j/\Delta, \label{lag_b}\\
\Delta&=&\frac{\partial (x_1,x_2,x_3)}{\partial (X_1,X_2,X_3)}. \label{lag_jac}
\eeqa
        Here $\B{x}=(x_1,x_2,x_3)$ is the current position of the fluid element 
	which is initially 
        at $\B{X}=(X_1,X_2,X_3)$, $\B{v}$ is the velocity moving with the fluid, 
        $\rho$ and $\rho_0$ are the current and initial densities of the fluid 
        element, $\B{B}=(B_1,B_2,B_3)$ and $\B{B}_0=(B_{01},B_{02},B_{03})$ are the
        current and initial magnetic fields and $P$ is the pressure. 

        Equations \ref{lag_mom} and \ref{lag_pos} are advanced in time using a
        Lax-Wendroff type method (fourth order in space, second order in time).
        Once the new positions of the fluid elements are known the other variables 
        can be calculated directly from equations \ref{lag_eng}-\ref{lag_jac} without
        further time integration. The method preserves total mass, entropy and 
        $\nabla \cdot \B{B}=0$ identically and gives excellent energy conservation.

        As time progresses the grid deforms moving more points into regions of
        compression. For the time evolution considered here, with the inner part
        of the twisted loop being forced by the instability into the almost static 
        external potential field, the grid points accumulate in the regions where
        the current sheet forms. This allows these increasing currents to be resolved
        for much larger values than would be possible by the equivalent Eulerian
        code. It should be noted, however, that as the Lagrangian code relies on the
        system being ideal it can say nothing about the evolution of the system 
        once dissipation becomes important.

{}

\newpage
 
\figcaption[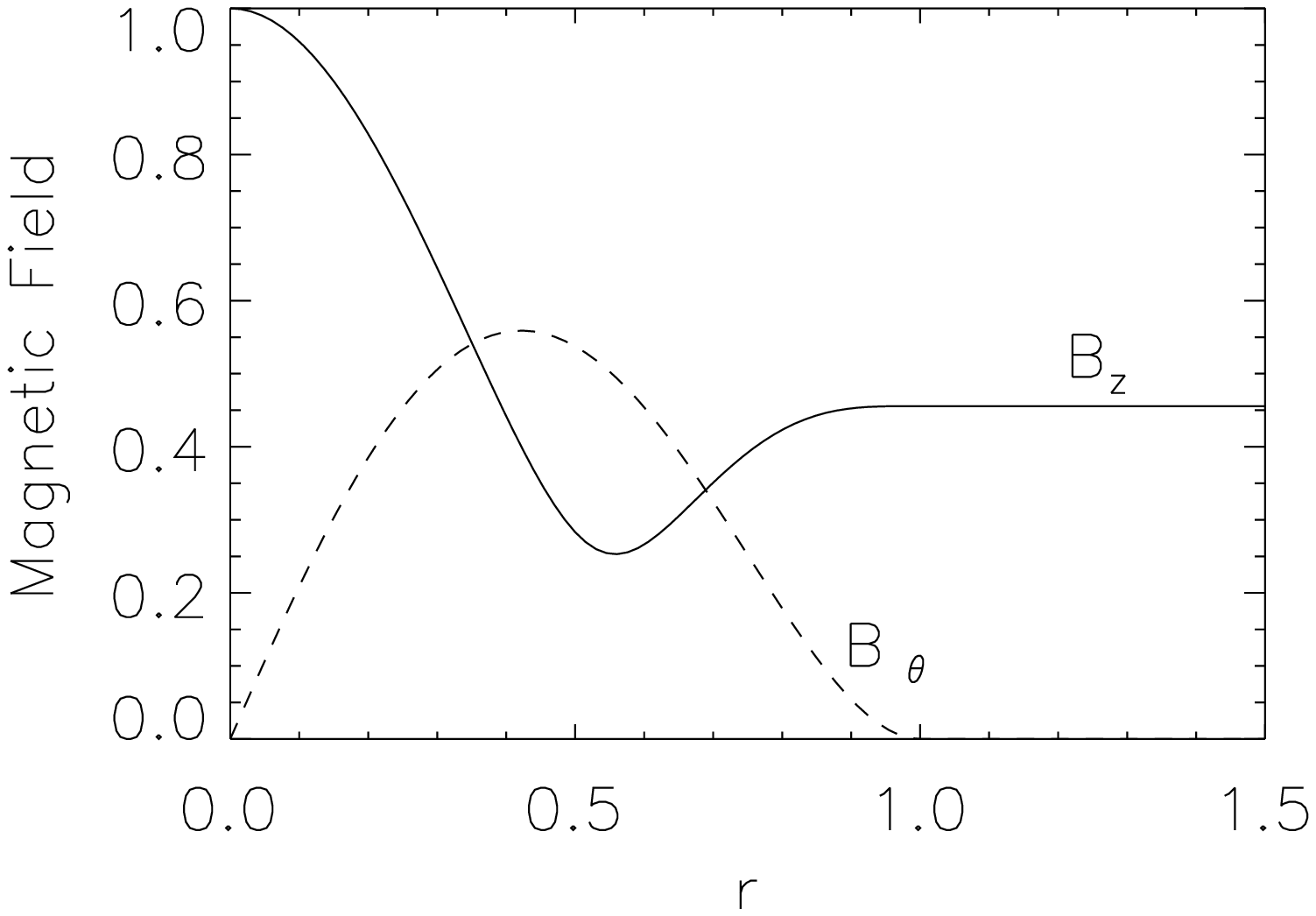]{Magnetic field components $B_{\theta}$ and $B_z$ vs.
        radius for the initial equilibrium. \label{fig1}}
\figcaption[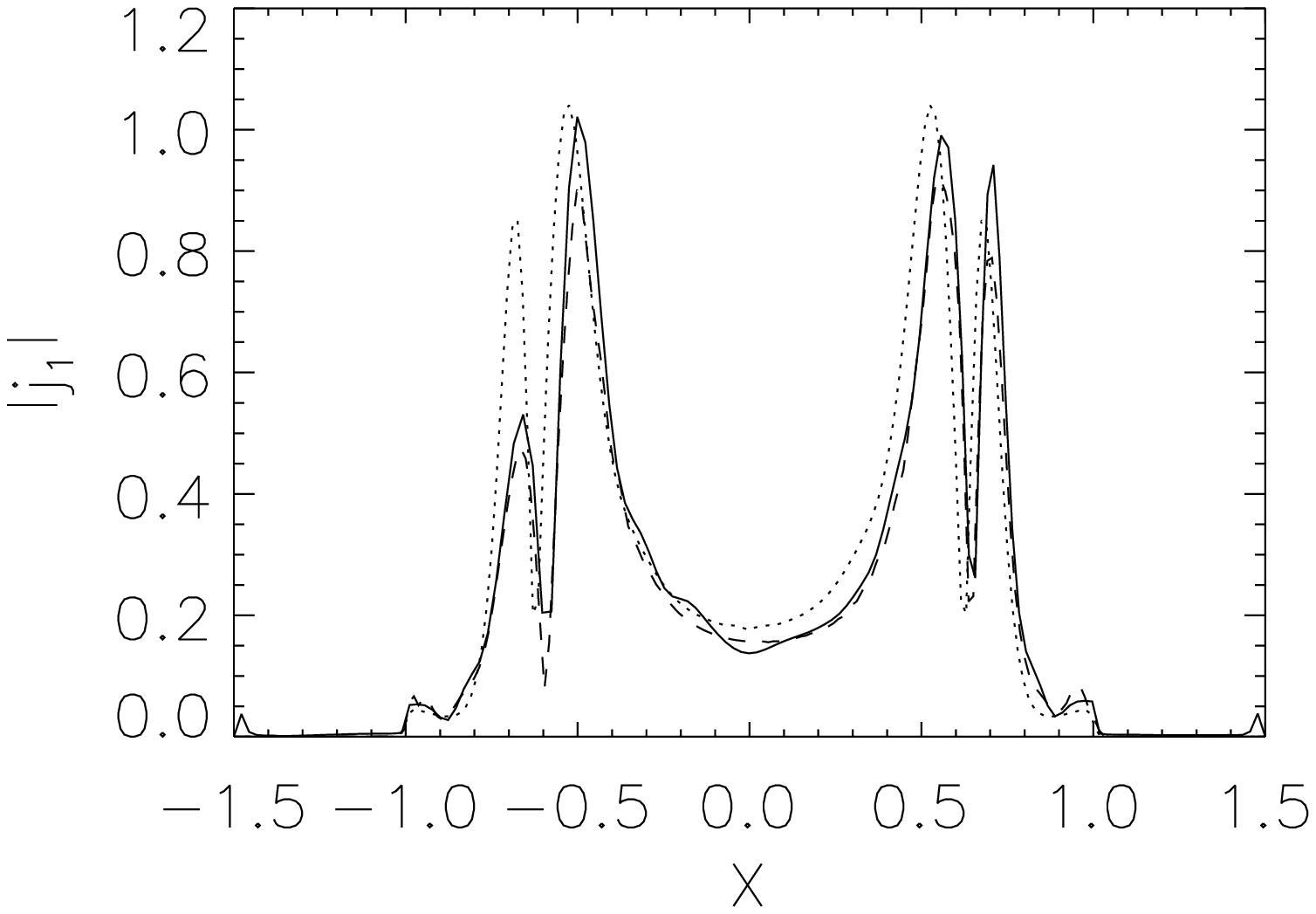]{The modulus of the current perturbation, $|\B{j}_{1}|$
        at $y=z=0$, $t=5$, from the linear (dotted line), the 
	non-linear Lagrangian (solid line) 
	and non-linear Eulerian (dashed line) time-evolution simulations. \label{fig2}}
\figcaption[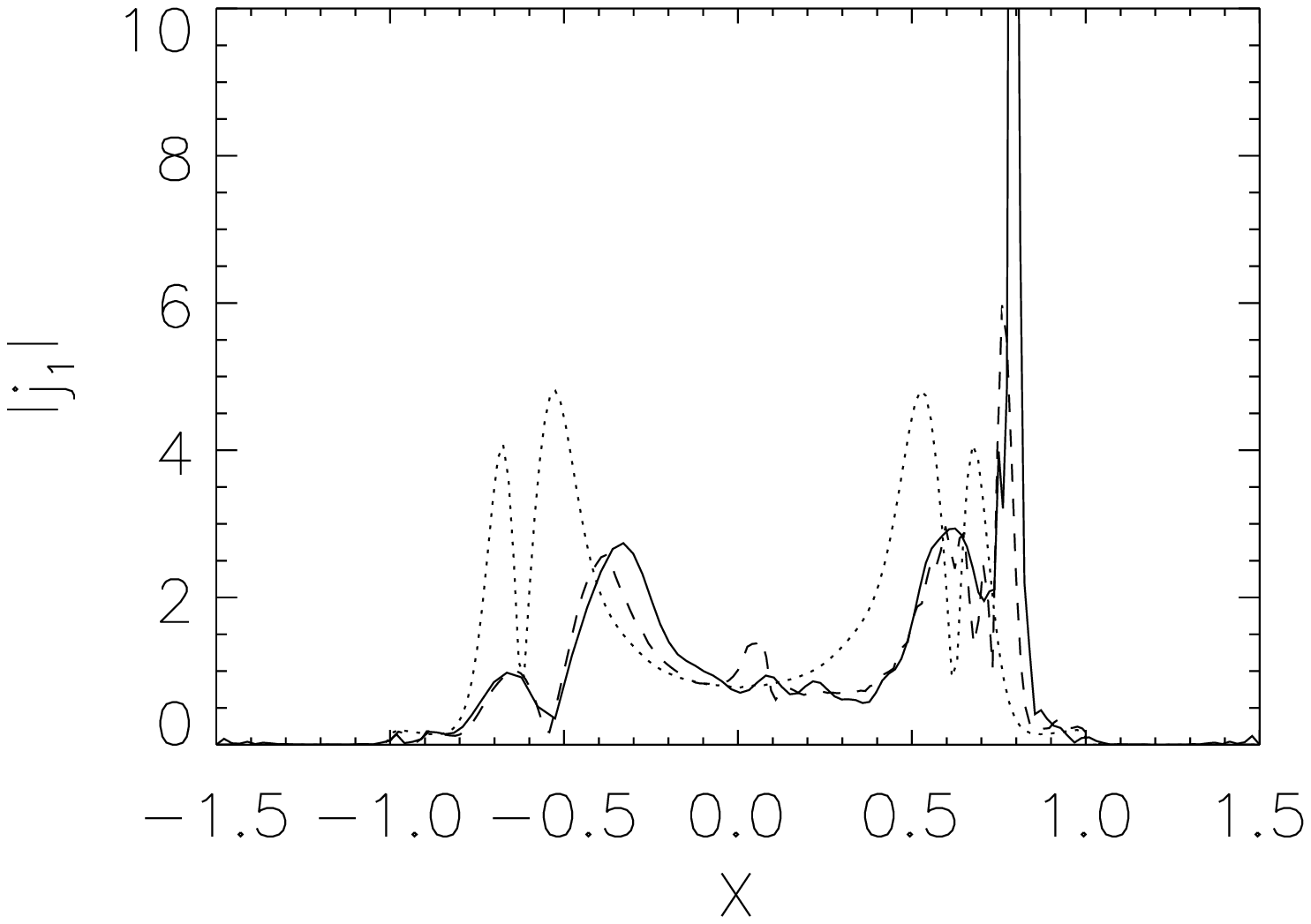]{As in Figure \protect\ref{fig2} but for $t=10$.
        \label{fig3}}
\figcaption[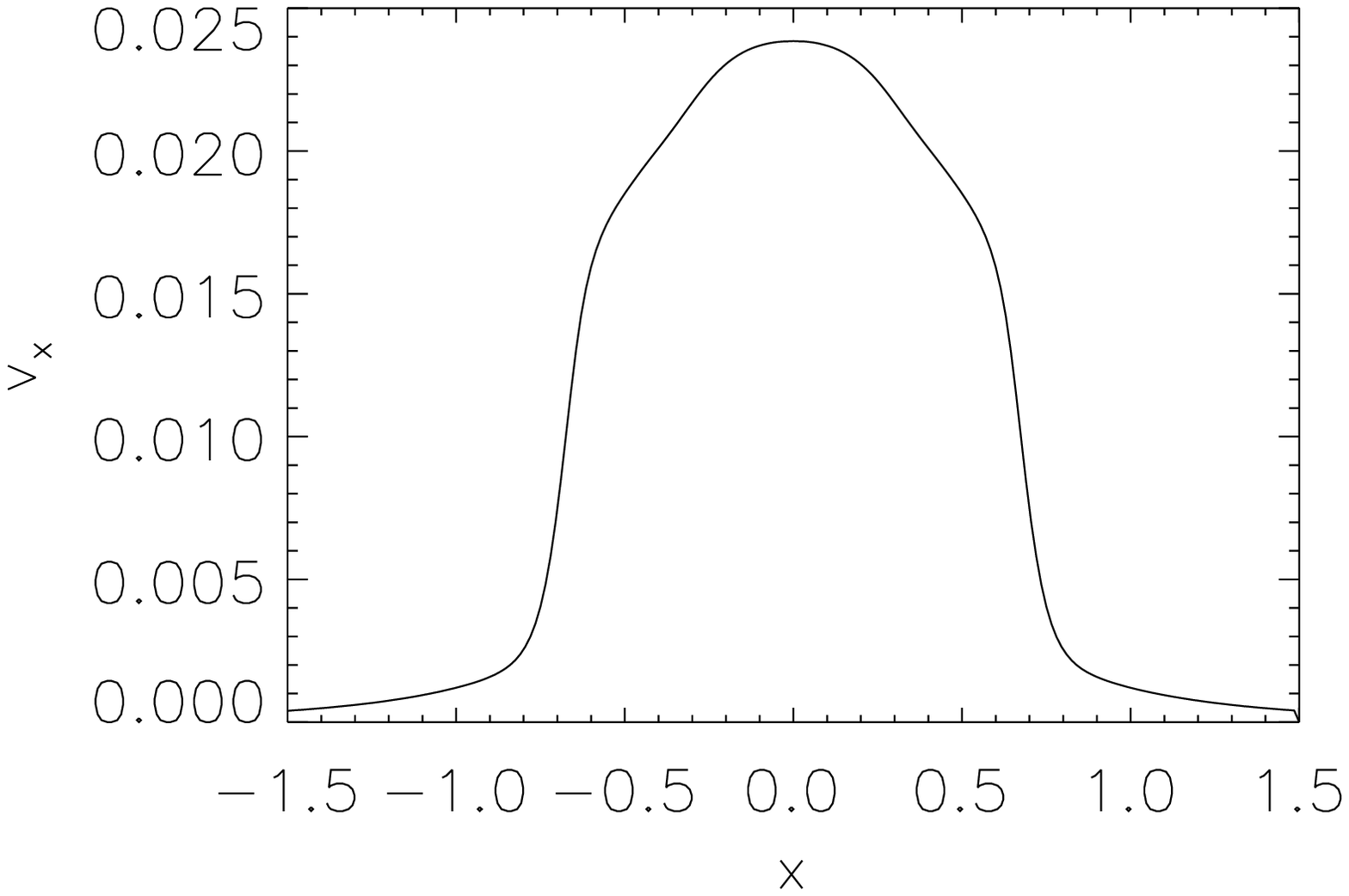]{The $x$-component of the velocity, $v_{x}$ at $y=z=0$,
        $t=5$, taken from the linear time evolution.
        \label{fig4}}
\figcaption[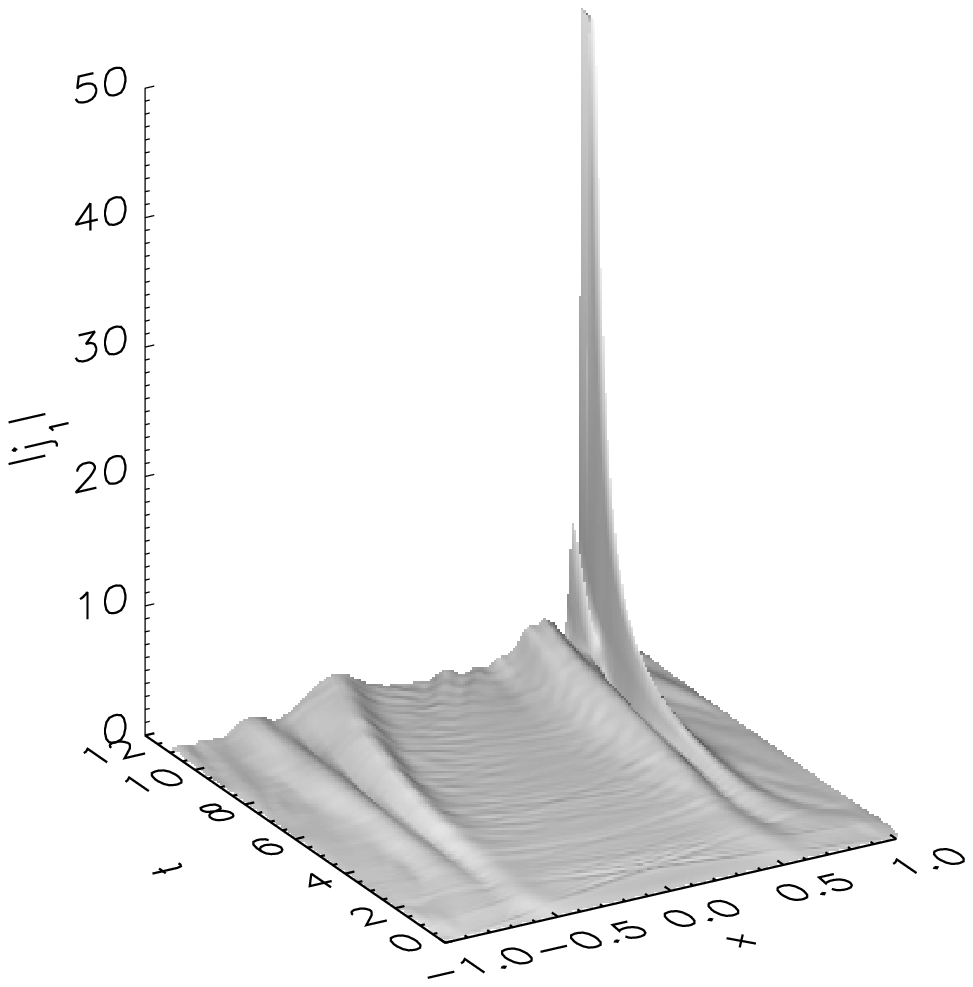]{The modulus of the current perturbation, $|\B{j}_{1}|$
        at $y=z=0$ as a function of $x$ and time from the nonlinear Lagrangian 
        simulation. \label{fig5}}
\figcaption[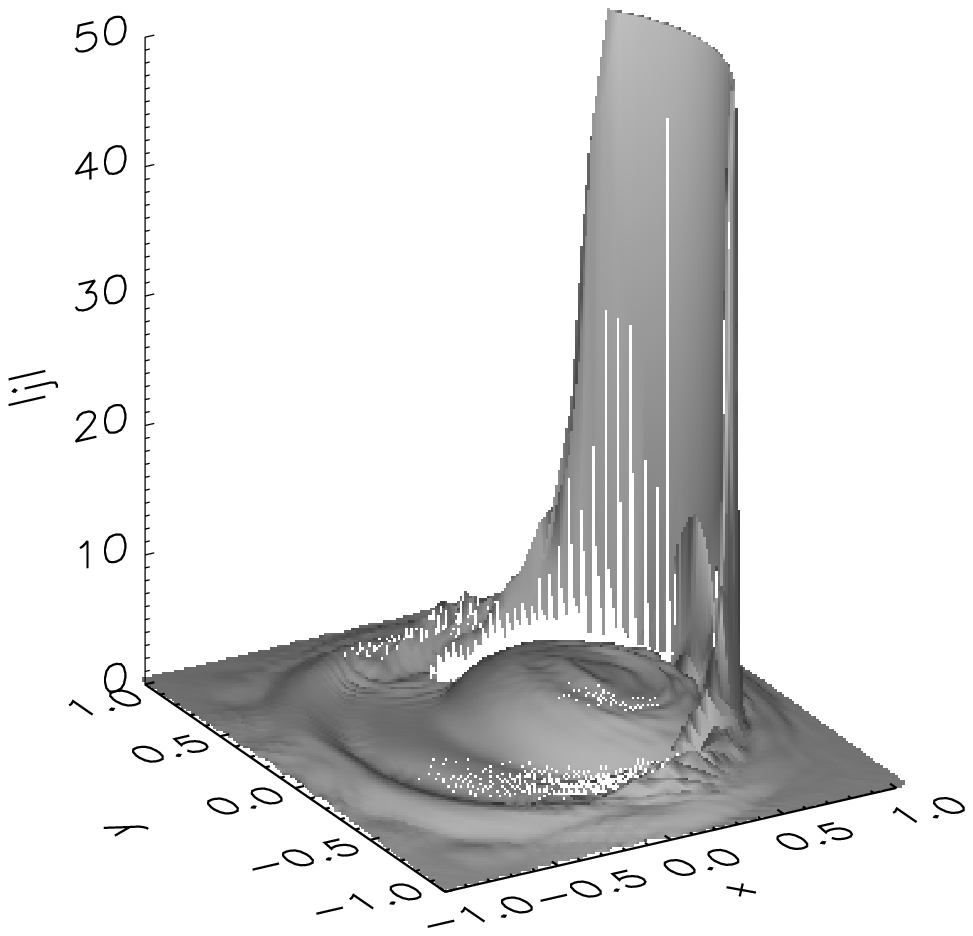]{The modulus of the current, $|\B{j}|$
        at $z=0$ from the nonlinear Lagrangian simulation at $t=11.5$. \label{fig6}}
\figcaption[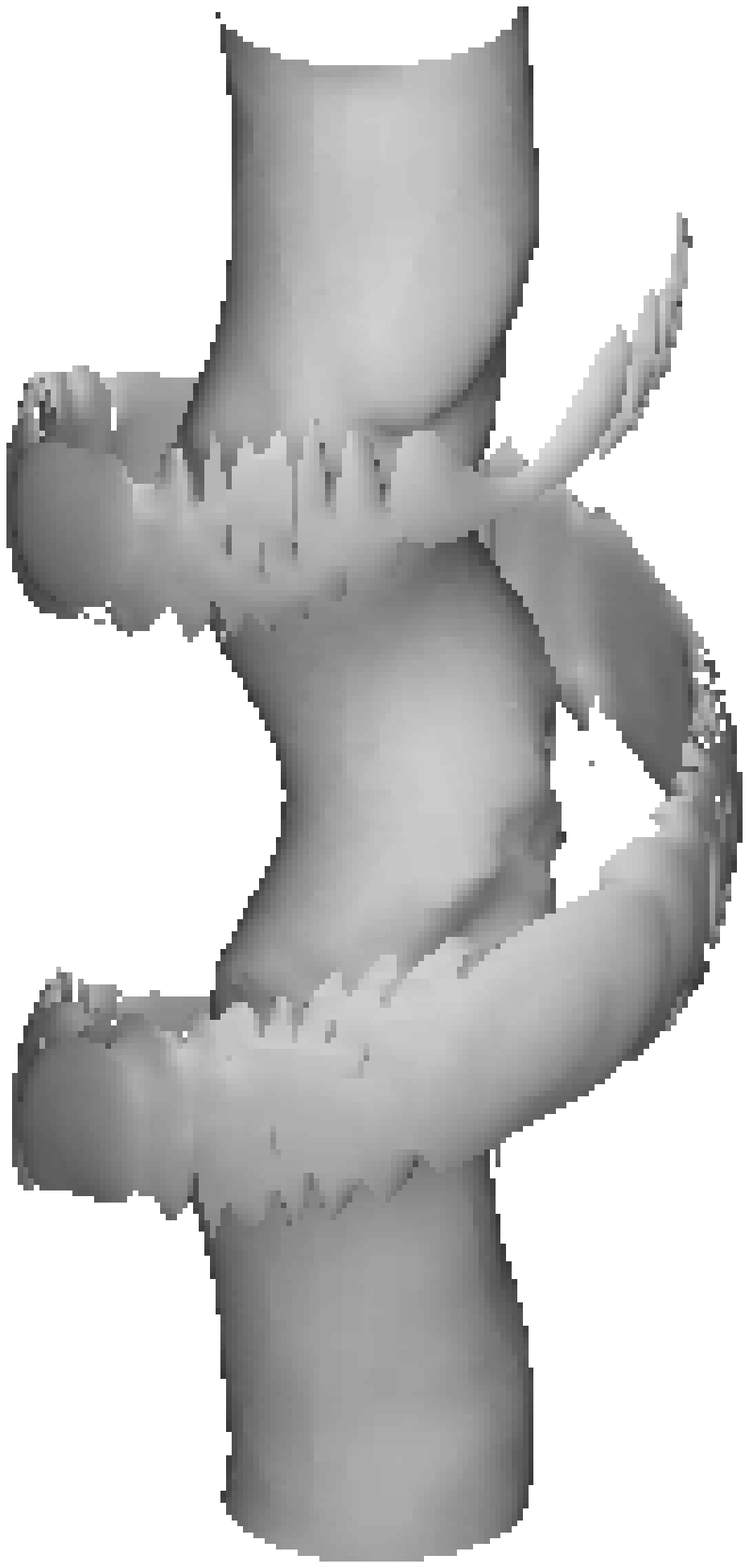,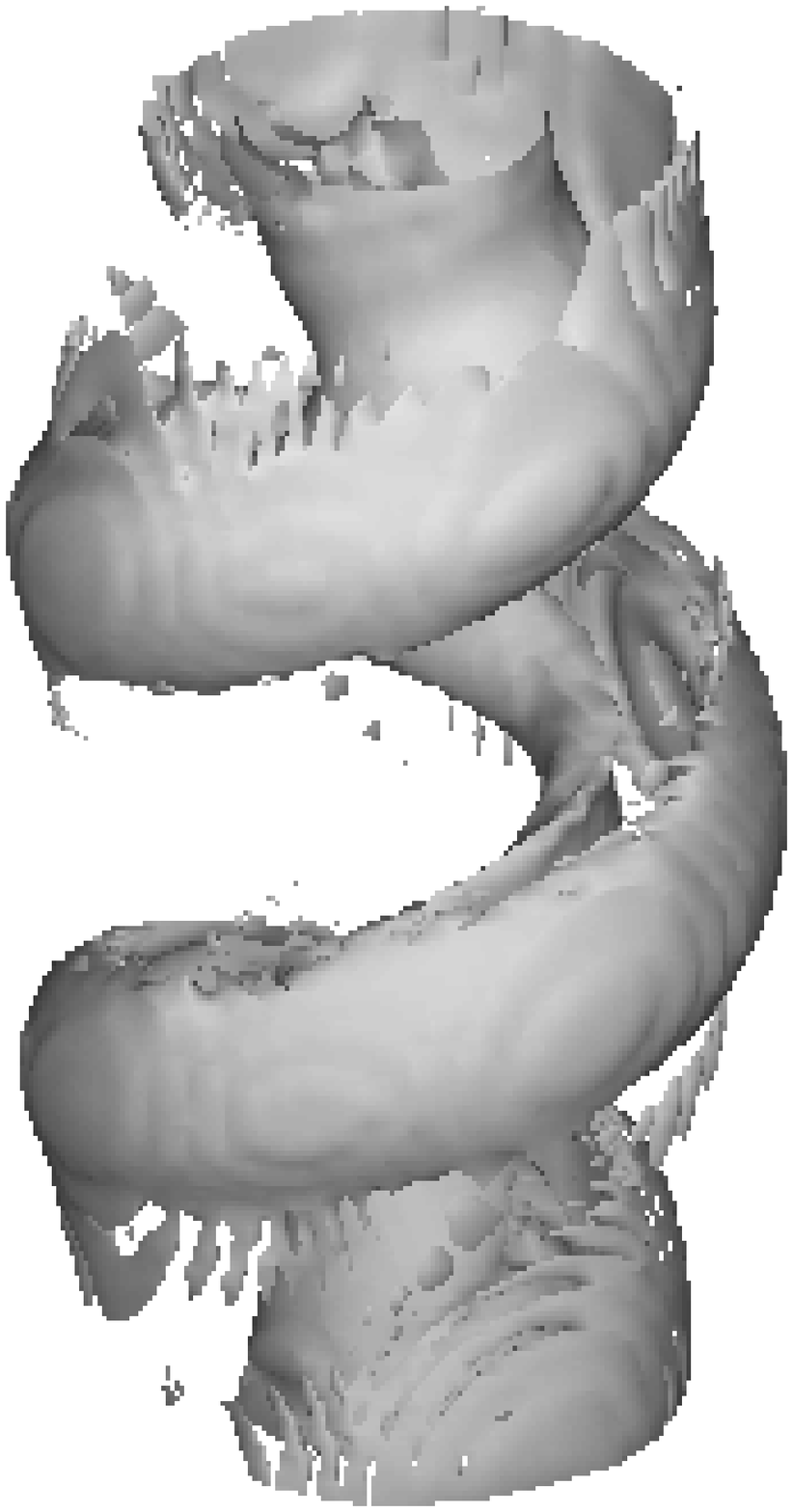]
	{Iso-surfaces of $|j|=3$ at $t=10$, and $t=15$ from the Eulerian
	simulation. \label{fig7}}
\figcaption[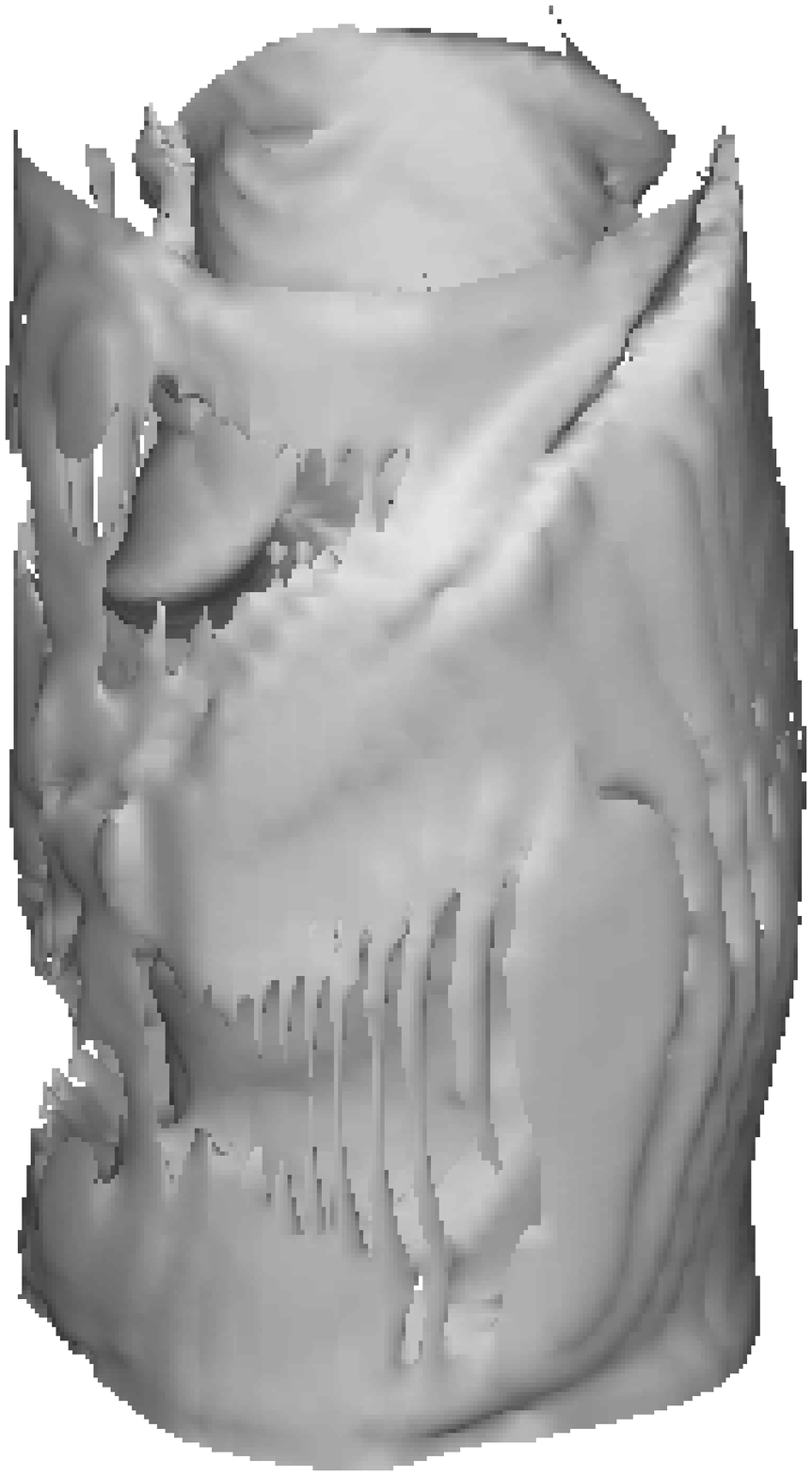,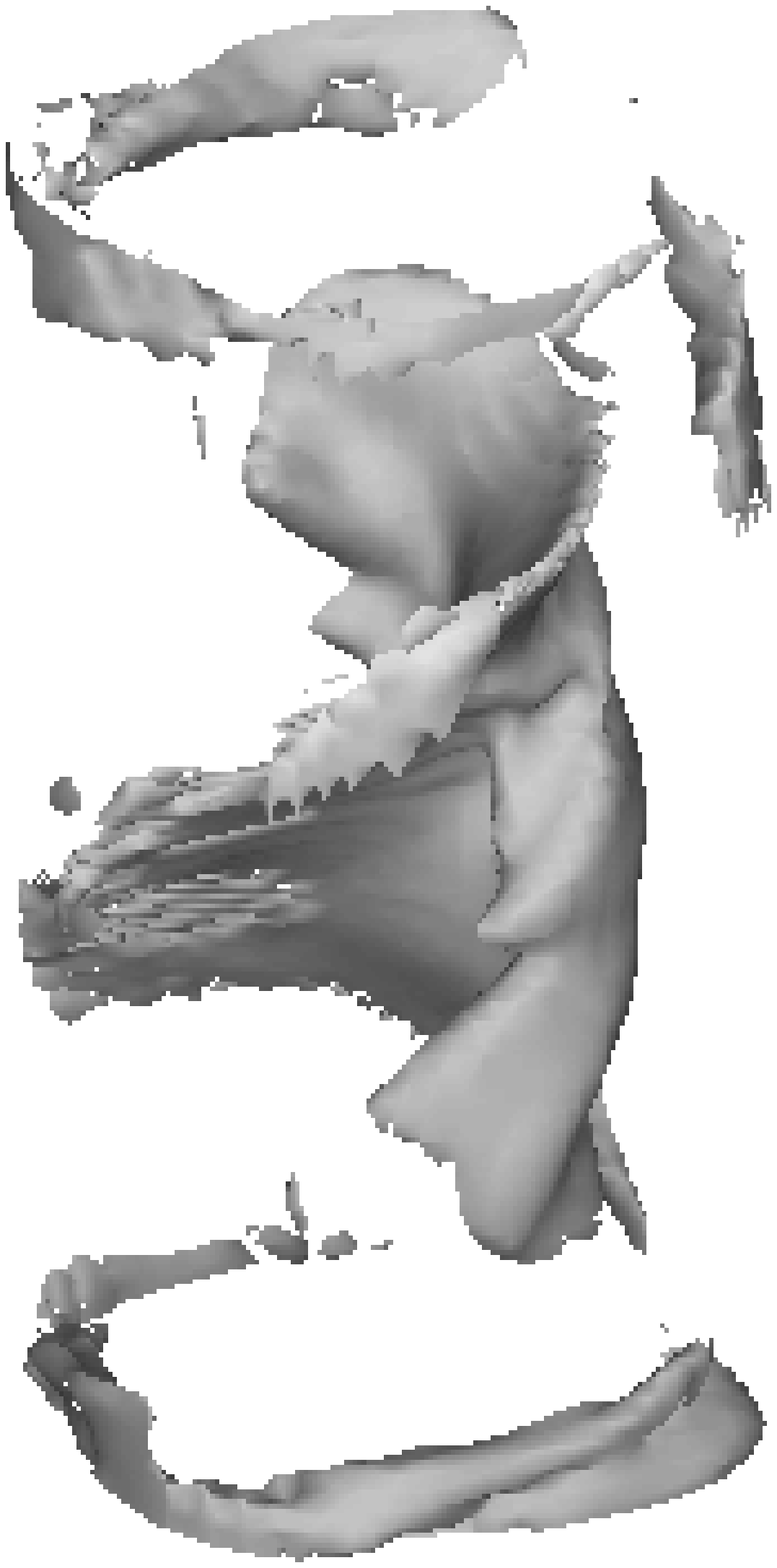]
	{Iso-surfaces of 3 times background energy and 6 times background at
	$t=20$ from the Eulerian simulation. \label{fig8}}
\figcaption[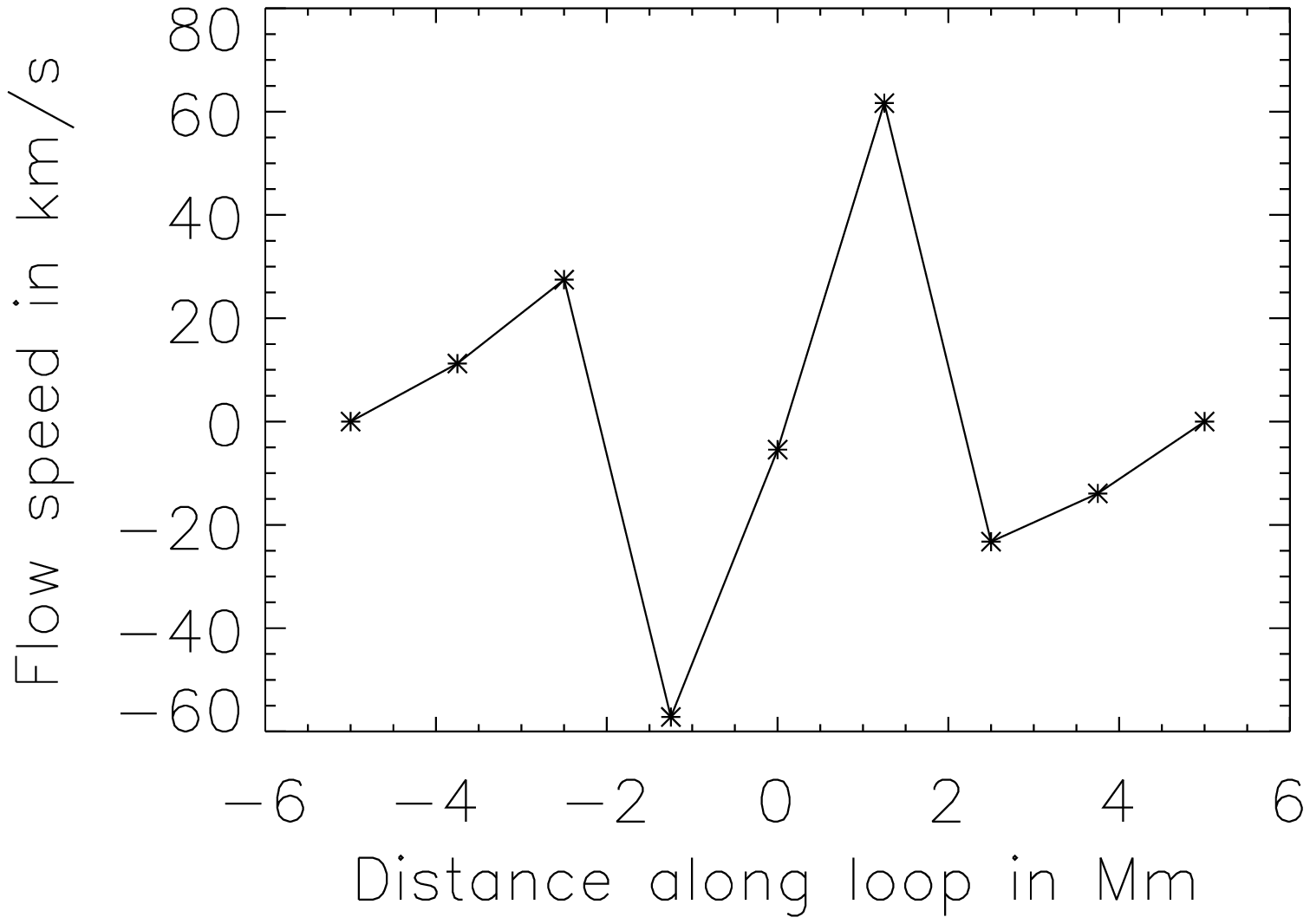]{Plasma velocity after averaging over a 10 seconds
	exposure time and a $1.5\,Mm \times 1.5\,Mm$ square vs. distance
	along the loop. Photospheric footpoints are at $\pm 5 Mm$. \label{fig9}}

\end{document}